\documentclass[11pt]{article}
\usepackage{amsmath}
\begin{document}
\title{Presence of exotic matter in the Cooperstock and Tieu galaxy model}
\author{D. Vogt\thanks{e-mail: danielvt@ifi.unicamp.br}\\
Instituto de F\'{\i}sica Gleb Wataghin, Universidade Estadual de Campinas\\
13083-970 Campinas, S.\ P.\ , Brazil
\and
P. S. Letelier\thanks{e-mail: letelier@ime.unicamp.br}\\
Departamento de Matem\'{a}tica Aplicada-IMECC, Universidade Estadual\\
de Campinas 13083-970 Campinas, S.\ P.\ , Brazil}
\maketitle
\begin{abstract}
We analyze the presence of an additional singular thin disk in the recent General Relativistic model of galactic gravitational 
field proposed by Cooperstock and Tieu. The physical variables of the disk's energy-momentum
tensor are calculated.  We show that the disk 
 is made of exotic matter, either   cosmic strings or
struts with negative energy density.
\end{abstract}
\section{Introduction}

Recently Cooperstock and Tieu \cite{Cooperstock} have proposed a General Relativistic
galactic model, which is essentially composed of a uniformly rotating fluid without pressure,
and fitted flat rotation curves without the (apparent) need of dark matter
with unusual
 properties. However, Korzy\'nski
\cite{Korzynski} argued that their model has an additional source of gravitational
field in the form of a rotating flat disk on the galactic plane, and thus should be considered 
``unphysical''. We think that the fact of having a flat disk is not a relevant
problem if the matter content of the  disk is made of usual matter. 
We agree with Korzy\'nski that some of the equations derived by Cooperstock and 
Tieu are not satisfied on the disk. 

We shall compute the matter content of an additional thin rotating disk in
 Cooperstock and Tieu's model, and  show that the disk is composed of exotic matter.
\section{Cooperstock and Tieu galaxy model}

In \cite{Cooperstock} the authors use the van Stockum metric \cite{Stockum}
\begin{equation} \label{eq_metric}
\mathrm{d}s^2=(\mathrm{d}t-N\mathrm{d}\varphi)^2-r^2\mathrm{d}\varphi^2-
e^{\nu}(\mathrm{d}r^2+\mathrm{d}z^2) \mbox{,}
\end{equation}
and propose a solution of the following form to fit the galactic rotation curves:
\begin{align}
\Phi &=\sum_{n} C_ne^{-k_n|z|}J_0(k_nr) \mbox{,} \\
N &=r \frac{\partial \Phi}{\partial r} \mbox{.}
\end{align}
The absolute value of $z$ is introduced to provide reflection symmetry of the 
matter distribution with respect to the plane $z=0$. Although the presence of $|z|$ 
does not modify the field equation for the density distribution
\begin{equation} \label{eq_rho} 
\rho = \frac{c^2}{8\pi G} \frac{N_{,r}^2+N_{,z}^2}{r^2} \mbox{,}
\end{equation} 
it introduces a distributional singularity at $z=0$ in the equation for the metric
function $N$:
\begin{equation} \label{eq_N}
N_{,rr}+N_{,zz}-\frac{N_{,r}}{r}=0 \mbox{,}
\end{equation}
and an additional distributional term in the energy-momentum tensor $T_{ab}$ \cite{Korzynski}. We shall now
calculate explicitly all the distributional components of $T_{ab}$.
\section{The van Stockum class of metrics}

For metric equation (\ref{eq_metric}), the exact Einstein field equations can be cast as \cite{Stockum}, \cite{Bonnor}
\begin{subequations}
\begin{align}
& N_{,rr} +N_{,zz}-\frac{N_{,r}}{r}=0 \mbox{,} \label{eq_stockum1} \\
& \nu_{,z} = -\frac{N_{,r}N_{,z}}{r}, \qquad \nu_{,r}=\frac{N_{,z}^2-N_{,r}^2}{2r} \mbox{,} \label{eq_stockum2}\\
& \rho =\frac{1}{r^2e^{\nu}}\left( N_{,r}^2+N_{,z}^2 \right) \mbox{,}
\end{align}
\end{subequations}
(we use units such that $c=8\pi G=1$). The introduction of an absolute value of $z$ in a solution 
of Eq.\ (\ref{eq_stockum1}) can be thought as doing a transformation $z \rightarrow |z|$ in the metric 
functions $N$ and $\nu$ (every transformation applied on $N$ will also affect the function $\nu$ by 
Eq.\ (\ref{eq_stockum2})). In the Einstein tensor we
have first and second derivatives of $z$. Since $\partial_z |z|=2 \vartheta(z)-1$ and $\partial_{zz} |z|=2\delta(z)$, where
$\vartheta(z)$ and $\delta(z)$ are, respectively, the Heaviside function and the
Dirac distribution, the Einstein equations yield an distributional energy-momentum tensor $T_{ab}=Q_{ab}\delta(z)$, with \cite{Gonzalez}
\begin{equation} \label{eq_emt}
Q^a_b=\frac{1}{2}\left[ b^{az}\delta^z_b-b^{zz}\delta^a_b+g^{az}b^z_b-g^{zz}b^a_b+b^c_c(g^{zz}\delta^a_b-g^{az}\delta^z_b) \right] \mbox{.}
\end{equation}
Here $b_{ab}$ denote the discontinuity functions of the first derivatives with respect of $z$ of the metric tensor on the plane $z=0$: 
\begin{equation} \label{eq_b}
b_{ab}= \left. g_{ab,z} \right|_{z=0^+}-\left. g_{ab,z} \right|_{z=0^-} =2\left. g_{ab,z} \right|_{z=0^+} \mbox{.}
\end{equation}
Using Eq.\ (\ref{eq_emt})--(\ref{eq_b}) and metric (\ref{eq_metric}), the only non-zero components of $Q^a_b$ are 
\begin{subequations}
\begin{align}
Q^t_t &=\frac{1}{e^{\nu}}\left( \frac{NN_{,z}}{r^2}-\nu_{,z} \right) \mbox{,} \label{eq_Qtt} \\
Q^t_{\varphi} &=-\frac{N_{,z}}{e^{\nu}} \left( 1+\frac{N^2}{r^2} \right) \mbox{,} \\
Q^{\varphi}_t &=\frac{N_{,z}}{r^2e^{\nu}} \mbox{,} \\
Q^{\varphi}_{\varphi} &=-\frac{1}{e^{\nu}}\left( \frac{NN_{,z}}{r^2}+\nu_{,z} \right) \mbox{.} \label{eq_Qphiphi}
\end{align}
\end{subequations}
The physical variables of the disk are obtained by solving the eigenvalue problem for $Q^a_b$: $Q^a_b\xi^b=\lambda\xi^a$, and has the 
solutions
\begin{align}
& \lambda_{\pm}=\frac{T}{2}\pm \frac{\sqrt{D}}{2}, \quad \text{where} \\
& T=Q^t_t+Q^{\varphi}_{\varphi}, \qquad D=(Q^t_t-Q^{\varphi}_{\varphi})^2+4Q^t_{\varphi}Q^{\varphi}_t \mbox{.}
\end{align}
Using Eq.\ (\ref{eq_Qtt})--(\ref{eq_Qphiphi}), we obtain
\begin{equation}
T=-\frac{2\nu_{,z}}{e^{\nu}}, \qquad D=-\frac{4N_{,z}^2}{r^2e^{2\nu}} \mbox{.}
\end{equation}
The discriminant $D<0$ characterizes heat flow in the tangential direction \cite{Gonzalez}, \cite{Moller}. In this case, the 
surface energy density and azimuthal stresses are, respectively, $\sigma=T/2$ and $P_{\varphi}=-T/2$. Thus, the disk
is composed of matter with an equation of state $P_{\varphi}=-\sigma$. If $\sigma>0$ we have tensions and it may 
be interpreted as an equation of state of matter formed by concentric
 loops of cosmic strings \cite{Letelier}. If $\sigma<0$ we also have exotic matter
with negative energy density. Objects with this equation of state are known in the literature as struts and they appear to stabilize 
certain superpositions of static isolated bodies in General Relativity (see, for instance, \cite{Bach}).
We also note that all the above results are exact.

Although the proposed galactic model does not really resolves galactic rotation without the presence of exotic matter, we 
believe that the idea of treating the non-linear galactic dynamical problem in the context of General Relativity is quite interesting
and should be further investigated, specially the rotating models where we have the non-Newtonian effect of dragging of inertial 
frames; a modest step in this direction is presented in \cite{Vogt}. 

\bigskip 
D. Vogt thanks CAPES for financial support. P. S. Letelier thanks CNPq and 
FAPESP for financial support.

\end{document}